\documentclass[11pt,fleqn,twoside]{article}

\newif\ifreport
\reporttrue

\usepackage[english]{babel}
\usepackage{derireport}

\usepackage{graphicx}
\usepackage[tight]{subfigure}
\usepackage{verbatim}
\usepackage{epsfig}
\usepackage[tbtags]{amsmath}
\usepackage{amsfonts}
\usepackage{amssymb}
\usepackage{url}
\usepackage{cite}
\usepackage{colortbl}
\usepackage{tabularx}
\usepackage[T1]{fontenc}
\usepackage{ae,aecompl}
\usepackage{pslatex}
\usepackage{pms}
\usepackage[leftcaption]{sidecap}
\usepackage{algorithm}
\usepackage{algpseudocode}
\usepackage{enumitem,lipsum}
\usepackage{flushend}

\usepackage{tikz}
\usetikzlibrary{fit,positioning}
\usepackage{tkz-graph}
\usetikzlibrary{arrows}

\def\ShowEdnotes{0}

\newcounter{myDefCounter}

\newenvironment{myDefinition}[1]
{\noindent\refstepcounter{myDefCounter}{\textit{Definition \arabic{myDefCounter}} (\textit{#1}).}\ }
{\hspace*{\fill}~\mbox{\rule[0pt]{1.3ex}{1.3ex}}\par\endtrivlist\unskip\medskip}

\foreignreport

\title{Virtual Location-Based Services: Merging the Physical and Virtual World}

\authornames{
Christian von der Weth
\and
Vinod Hegde
\and
Manfred Hauswirth
}

\innertitle{Virtual Location-Based Services:\\Merging the Physical and Virtual World}

\author{
Christian von der Weth%
\affiliation{
DERI (Digital Enterprise Research Institute), National University of Ireland,
Galway\protect, IDA Business Park, Lower Dangan, Galway, Ireland.
\mbox{E-mail: christian.vonderweth@deri.org.}}
\and
Vinod Hegde%
\affiliation{
DERI (Digital Enterprise Research Institute), National University of Ireland,
Galway\protect, IDA Business Park, Lower Dangan, Galway, Ireland.
\mbox{E-mail: vinod.hegde@deri.org.}}
\and
Manfred Hauswirth%
\affiliation{
DERI (Digital Enterprise Research Institute), National University of Ireland,
Galway\protect, IDA Business Park, Lower Dangan, Galway, Ireland.
\mbox{E-mail: manfred.hauswirth@deri.org.}}
}

\published{
}

\acknowledgement{
This work has been funded (in a part) by
    Science Foundation Ireland (Grant no.~SFI/08/CE/I1380 -- L\'{i}on-2)
}

\abstract{
Location-based services gained much popularity through providing users with helpful information with respect to their current location. The search and recommendation of nearby locations or places, and the navigation to a specific location are some of the most prominent location-based services. As a recent trend, virtual location-based services consider webpages or sites associated with a location as 'virtual locations' that online users can visit in spite of not being physically present at the location. The presence of links between virtual locations and the corresponding physical locations (e.g., geo-location information of a restaurant linked to its website), allows for novel types of services and applications which constitute virtual location-based services (VLBS). The quality and potential benefits of such services largely depends on the existence of websites referring to physical locations. In this paper, we investigate the usefulness of linking virtual and physical locations. For this, we analyze the 
presence and distribution of virtual locations, i.e., websites referring to places, for two Irish cities. Using simulated tracks based on a user movement model, we investigate how mobile users move through the Web as virtual space. Our results show that virtual locations are omnipresent in urban areas, and that the situation that a user is close to even several such locations at any time is rather the normal case instead of the exception.
\\*[\parskip]
~
\\*[\parskip]
}

\deritr{2013-10-10}
\date{October 2013}

\begin{document}

%
%

\maketitle

\newpage 
\pagenumbering{Roman}

{\small

\tableofcontents
}

\cleardoublepage

\section{Introduction}
\label{sec:introduction}
With the advances in mobile technologies, people can go online and access the Web almost everywhere. Modern mobile devices have sophisticated sensors on board which can be used to determine the physical context of a user such as current location, speed of movement, etc. These developments have spurred the success and popularity of a class of mobile applications which provide mobile users with information based on their physical context. The location of users is one ofthe most important physical context as evident by the huge success of location-based services such as Foursquare, Yelp, Google Places etc. Common types of location-based services include: 
(a) identifying a user's location and providing information about it;
(b) searching for locations or users within an area;
(c) navigating users to a specific location; and
(d) communicating, socializing or collaborating with other users nearby.

In previous works~\cite{vdw11COBS,vdw11VirtualPresence}, we motivated and studied the advantages of linking locations with the corresponding websites as basic idea behind the notion of virtual location-based services (VLBS). In a nutshell, VLBS consider an individual webpage, sets of pages or complete websites representing a location as a \textit{virtual location}. For example, given a location like the computer science building of a university, the website for that computer science department denotes a virtual location. It is easy to see that the virtual locations a user visits are good indicators of the user's interests or information needs. Thus, users browsing the same page share either common interests or look for the same information. VLBS utilize this idea to provide additional information about (virtual) locations or enable them to communicate and collaborate between users physically on-site. In~\cite{vdw11VirtualPresence}, we furthermore aim to merge both types of location-based services by 
connecting physical with virtual locations, in an overarching model of ``space''. 

In this paper, we investigate the feasibility, usefulness and potential benefits of VLBS in detail. The success of VLBS depends on the widespread existence of links between locations and their virtual locations. To the best of our knowledge, a quantitative evaluation of links between locations and their virtual locations is still missing. More specifically, we formulate the following research questions:
\begin{enumerate}[label=(\arabic*)]
\item How common are locations with a physical as well as a virtual representation and how are they distributed across cities?
\item How do mobile users in the real world move through the virtual space in terms of being close to virtual locations?
\item What are new insights into the realization of VLBS to improve the online experience of both web and mobile users?
\end{enumerate}

To answer these questions, we provide an in-depth case study analyzing the existence and distribution of virtual locations. We focus on locations which ``naturally'' feature a physical counterpart. This includes websites dedicated to physical locations, e.g., the websites of shops, hotels, bars, parks, tourist attractions, etc. Also company and business websites typically refer to specific physical locations. VLBS exploit links between these locations and their virtual locations to provide novel functionalities to users such as context-aware interactions or computer-supported cooperative work on the Web. For example, a web user sitting at home and browsing a restaurant's website might be able to contact mobile users present in the restaurant to inquire about the number of people in the restaurant (geo-social search). A mobile user walking by a shop can be notified about the shop's website to provide the user with interesting information about current offers in the shop (mobile web advertising).
 Also new ideas for mobile gaming involving both web and mobile users based on VLBS are conceivable.

For our case study, we collected virtual locations within the cities of Dublin and Galway, Ireland. Both cities differ significantly regarding the geographic and population size, allowing us to investigate the effects on the existence and distribution of virtual locations. Furthermore, we use the real-world user visits on Foursquare\footnote{https://foursquare.com/} to simulate user movements in both cities in order to investigate mobile users' visits at virtual location while moving through virtual space. We distinguish between two types of movements: 
\textit{recurring movements} refer to users' movements that are part of their daily routine such as going to work;
\textit{non-recurring movements} refer to less common movements, e.g., going to a pub after watching a movie in a movie theatre on the weekend.
Our analysis of the location datasets shows that in urban environments virtual locations are virtually everywhere with high peaks in the city centers, commercial districts or touristic areas. The results of our simulation of user movements show that most of the physical user movements within a city result in them traversing through many intermediate virtual locations. This is particularly true for non-recurring movements, since here mobile users are more likely to pass areas with a high density of virtual locations. These results indicate that effective and useful VLBS can be developed for urban areas.

The remainder of the paper is organized as follows: 
Section~\ref{sec:relatedwork} reviews related work to put our approach into context. 
Section~\ref{sec:vlbs} provides a basic model of the virtual space and presents our proof-of-concept implementation of a VLBS application. 
Section~\ref{sec:setup} presents our experiment setup by describing the involved data collection and generation process. 
Section~\ref{sec:evaluation} presents the results of our comprehensive case study.
Section~\ref{sec:roadmap} outlines challenges and open research questions towards the implementation of real-world VLBS.
Finally, Section~\ref{sec:conclusions} concludes and outlines on-going work.

\section{Related Work}
\label{sec:relatedwork}
The notion of location-based services generally covers a broad spectrum of research areas which include both technical and social aspects. In the following, we give an overview of the most relevant research efforts in this field.
\\
\\
\textbf{Location-based services.} Over the last decade, location-based services gained enormous popularity since mobile devices enabled users to get contextualized information based on their location ~\cite{Roza03OverviewOfLBS,Bellavista08LBS}. Common applications of location-based services include the navigation of users to a destination or a point of interest, the provisioning of information about a user's current location, the search for nearby events or points of interest, and others. The most basic challenge is the accurate calculation or estimation of the locations of mobile devices. For this, various location estimation techniques exist~\cite{Liu07SurveyOfWireless}. While GPS or cellular network-based solutions are used for outdoor location estimations, Wi-Fi or Bluetooth have been used to determine indoor locations of mobile devices ~\cite{Lassabe09IndoorPositioning,Gu09SurveyOfIndoor}. The exact locations of users are considered as sensitive information which users are typically 
not willing to share. Existing approaches to preserve users' privacy aim to not disclose one's exact location but rather an estimate~\cite{Krumm09Survey,Shin12PrivacyProtection}. The main challenge here is identifying a meaningful trade-off between the level of privacy and the quality of the provided location-based service. Various user studies have been conducted to investigate users' preferences with respect to sharing their location with others, e.g.,~\cite{Brush10ExploringEndUser,Toch10EmpericalModels}. Besides privacy, other user studies such as~\cite{Chang07UserStudyOnAdoption,Junglas05ExperimentalInvestigation} investigate the effects of multiple factors (e.g., costs, security, quality) on the successful adoption of location-based services.
\\
\\
\textbf{Location-based social networks.} Location-based social networks enable users to establish social connections with others and express their visits to places along with their social profiles.  The services provided by location-based social networks such as checking-in at places, rating them and commenting about places are more sophisticated and user-centric as they also bring in their social context into consideration. Location-based social network platforms such as Facebook Places, Foursquare, and Google Places are some of the most popular ones. There have been various research efforts in analyzing the user visits to places and the effect of social ties between users on the user movement patterns \cite{cho2011friendship,scellato2011socio,zheng2010geolife}. The results in \cite{backstrom2010find} indicate that social ties of users can be used to discover approximate locations of users. \cite{wang2011human,scellato2011exploiting} show that user mobility patterns can be used to predict the social ties between users. The mentioned research efforts show there exists a strong link between social network of a user and his/her movements.
\\
\\
\textbf{Towards virtual location-based services.} The concept of virtual locations originates from the efforts towards collaboratively browsing and searching the Web. \textsc{SearchTogether}~\cite{Morris07SearchTogether} and \textsc{CoScripter}~\cite{Leshed08CoScripter} enable collaborative browsing between users working with their own computers. They allow any group of users to initiate joint browsing on a website, recognizing that many tasks and information needs on the Web demand the collaboration between users to maximize user benefits. \textsc{PlayByPlay}~\cite{Wiltse09PlayByPlay} describes the need for collaborative browsing platforms for easily and efficiently browsing the Web. It also demonstrates the use of collaborative browsing with a system which lets the mobile device users and web users collaborate and communicate. \textsc{COBS} (\textit{CO}llaborative \textit{B}rowsing and \textit{S}earching)~\cite{vdw11COBS,
vdw11FAST} proposes a browser extension that provides users with features to directly to indirectly communicate and collaborate by performing tasks like adding tags, comments to websites. The extension~\cite{vdw10COBSdemo} provides a proof-of-concept implementation that allows users visiting the same website i.e., virtual location, to communicate with each other. All these works focus on providing services to only those users visiting virtual locations.  In~\cite{vdw11VirtualPresence}, we present a novel approach to enable the communication  between users  visiting virtual locations and users present at physical locations. In this work, we consider those virtual locations or websites which have been linked to their corresponding physical locations. We also implemented a web browser extension and a mobile application in order to demonstrate the possibility of instant communication between users at physical and virtual locations.
\\
\\
In summary, traditional location-based services and related services on the Web have been investigated independently. In~\cite{vdw11VirtualPresence}, we have proposed a framework to link the physical locations to their virtual locations in order to enable better communication and collaboration between users.  We have also presented some preliminary results regarding the potential benefits of the framework. In this paper, we show the usefulness for merging physical and virtual locations in order to develop novel virtual location-based services. We use openly available datasets for two cities in Ireland which have many of their physical locations linked to the virtual locations.

\section{Virtual Location-Based Services}
\label{sec:vlbs}
Merging physical and virtual world in the ways we envision via VLBS is a rather novel idea. In this section, therefore, we introduce the notion of VLBS as follows: We first provide a model for the virtual space by defining the involved concepts. We then present VLIMSy, our current proof-of-concept implementation of a VLBS application, to showcase the potential benefits of such services.

\subsection{A Model of the Virtual Space}
\label{subsec:model}
Our approach is to adopt the notion of a user's location from the real world to the Web. In the following, we first define the required concepts of a virtual coordinate and virtual location. We limit the presentation of the model of the virtual space to the concepts required for this paper; we present the full model in~\cite{vdw11VirtualPresence}. We then outline the differences between the physical and virtual space, and discuss their effects on the design of virtual location-based services.
\\
\\
\textbf{Virtual coordinates and locations.}
Simply speaking, space describes the possibilities where a person ``can be''. Given these notions, we can define the virtual space as the set of web pages a user can visit. In geographic terms, the most fine-grained way to specify a mobile user's current position is by means geo coordinates, e.g., longitude and latitude. Mapping this concept to the virtual space, the current position of a user is the web page the user is visiting. Thus, within our framework, each page on the Web is uniquely identified by a URL.
\\
\\
\begin{myDefinition}{Virtual coordinate}
A virtual coordinate $vc$ is the URL of a webpage.
\end{myDefinition}

In many application contexts, not the distinct page but the category or topic or similar concepts of a page are of interest to describe a web user's location. We therefore extend the definition of a virtual location beyond a single virtual coordinate.
\\
\\
\begin{myDefinition}{Virtual location}
A virtual location $vl$ is a distinct, non-empty, finite set of virtual coordinates $\mathcal{V} = \{vc_1, vc_2, ..., vc_n\}$, with $\mathcal{V}_1 \cap \mathcal{V}_2 = \emptyset$.
\end{myDefinition}
The set of virtual coordinates that constitutes a virtual location is application-specific. Throughout this paper, we use the domain of a URL as identifier of a virtual location, i.e., we group all subpages of a website into one location. This is a reasonable assumption for websites associated to physical locations such as hotels, shops, businesses, etc., which are in the scope of our evaluation.
\\
\\
\textbf{Physical vs. virtual Space.}
\label{sec:physical_vs_virtual}
Despite sharing similar notions, physical and virtual locations have fundamental differences affecting the design and implementation of VLBS. 

\textit{Distance and locality.}  In the physical space, the distance between two locations is well-defined, e.g., using the Euclidean distance. Between two web pages, in general, such distance measures are missing. While one can define the distance between two pages $u_1$ and $u_2$, e.g., as the minimum number of hyperlinks needed to be followed to get from $u_1$ to $u_2$, this does not necessarily constitute a meaningful distance definition in an application context.  For our model of the virtual space, we utilize existing efforts to quantify the similarity between web pages~\cite{vdw11VirtualPresence}.

\textit{Moving between locations.} 
Natural limitations regarding the time required to move between physical locations as well as regarding possible directions often allow, to some extent, to predict a person's or object's location in the near future. On the Web, a user can navigate at any point in time to any page. Thus, to reliably predict the next page a user will navigate to is, in general, not possible. In practice, however, various heuristics to identify a set of virtual locations a user is likely to visit next may be applicable, e.g., if web users may show distinct patterns in their browsing behavior.

\textit{Identity protection.} A mobile user's location in the physical space is typically known to everybody in the user's vicinity. On the Web, however, mechanisms to shield or hide one's personal data are omnipresent. In social networks, for example, users create explicit connections to others and organize them into groups. Applied to VLBS, such mechanisms enable users to reveal their current location only to selected users.

\textit{Symmetry.}  In the physical space, in general, if a Person $A$ is aware of a Person $B$, so is $B$ of $A$. Mechanisms supporting to shield one's virtual location from others break this symmetry. Due to privacy concerns, web users have a strong incentive to hide, potentially resulting in a majority of hiding web users. Obviously, there is a trade-off between privacy protection mechanisms and incentives for users to actively contribute, i.e., share their current virtual location, provide user-generated content, etc.

\textit{Multiple locations.}  At any time, a mobile user in the physical space always occupies one unique location. On the Web, a user can visit multiple web pages by using multiple browser windows or tabs. Thus, a web user may be at different virtual locations and as such is aware of distinct groups of other web or mobile users. Any virtual location-based service towards group collaboration, e.g., chat messaging, have to distinguish between such groups.

\subsection{A Simple VLBS Application}
\label{subsec:implementation}
In this section, we describe VLIMSy (Virtual Location Instant Messaging Service), our proof-of-concept implementation of a simple virtual location-based service. In a nutshell, VLIMSy allows the exchange of presence information and messages of web and mobile users based on their physical and/or virtual location. For example, a web user browsing a shop's website can connect with other web user visiting the same site or mobile users close to the shop.
\begin{figure*}
  \centering
  \includegraphics[width=1.0\textwidth]{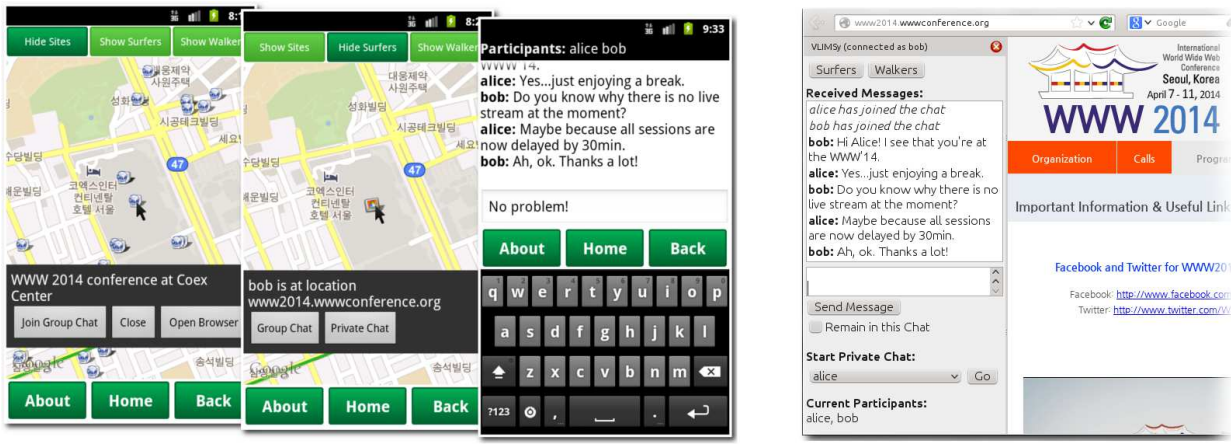}
  \caption{Screenshots of the frontend applications to showcase a typical use case covered by VLIMSy: User Bob is browsing the WWW'14 website and spotting conference attendee Alice. Now Bob can directly contact Alice to ask her for up-to-date information which is not available on the website.}
  \label{fig:vlimsy}
\end{figure*}
\\
\\
\textbf{Backend Architecture.}
A data repository maintains the mapping between the physical and virtual locations. We represent physical locations using 2-dim\-en\-sio\-nal geo coordinates, i.e., latitude and longitude.  For the time being, we focus on ``single point'' locations like hotels, pubs, shops, etc., without considering the spatial geometry. We also store locations with a large spatial extent, like parks or golf courses, using single geo coordinates to represent their physical location. We collected the current data in our repository using the \textsc{Google Places API} (see Section~\ref{sec:setup} for details). 

We provide the presence and instant messaging service based on the popular, open-standard eXtensible Messaging and Presence Protocol (XMPP\footnote{http://xmpp.org}). The protocol supports user-to-user chats as well group chats. Our backend features an XMPP server as the core component which enables any third-party solutions (e.g., instant messaging clients) with XMPP support to connect. For VLIMSy, the most relevant concept is the ``group chat``. We assign each location to a group chat. The intuition is that users at the same location are in the same group chat and are therefore aware of each other. In general, the physical and virtual locations of a user differ. For example, a customer in a shop is not necessarily browsing the shop's website. We therefore distinguish between a \textit{geo group chat} representing the physical representation and a \textit{web group chat} representing virtual representation of a location. Besides presence information, we also make use of the possibility to exchange messages. 
We support group chats and the user-to-user communication between web and mobile users.
\\
\\
\textbf{Frontend Applications.}
Given the different devices and applications for web users (at home or work) and mobile users, we provide two different interfaces for both user roles to interact with VLIMSy: a web browser add-on and a mobile application. Figure~\ref{fig:vlimsy} shows screenshots of the VLIMSy mobile application and the browser add-on, showcasing a typical interaction between a web and mobile user.

\textit{Web browser add-on.} 
From a usability perspective, we aim for a seamless integration of our presence mechanisms into the normal browsing experience of users. We therefore implemented a browser add-on featuring a sidebar to provide our instant messaging service in an unobtrusive manner. The add-on maintains an XMPP connection with references to two group chats, (a) to the web group chat of the currently visited website, and (b) to the geo group chat of the corresponding physical location (if available). The latter enables web users to be aware of mobile users that are close to the physical location of the visited website. The information window for the web group chat also allows users to communicate in a chat-like fashion: Users can send private messages to individual users or can send public messages to the group chat which can be read by all its current participants. 

\textit{Mobile phone application.} For mobile users, we implemented an Android application with two main features: The first one is a map based on the \textsc{Google Maps} API which displays all available virtual locations, web and mobile users in the vicinity as different markers. Clicking on a marker displays some basic information about the corresponding location, web or mobile user. This information window also includes buttons that allow a mobile user (a) to enter the group chat of a virtual location or (b) send a private message to other web or mobile users. The second feature is a basic chat client for private and group chats. We assume that a mobile device is capable of determining its current location. Every time the physical location changes, the application sends a request to the backend with the new geo coordinates. The response is the closest virtual location in a radius of, e.g., 100m. If such a location exists, the application automatically enters the corresponding geo group chat, and thus 
making mobile users visible to web users. 
\\
\\
VLIMSy is our current experimental setup to illustrate the potential of VLBS and to get deeper insights into the challenges of their implementation as real-world applications. In principle, our setup can easily be extended to meet the needs of users such as setting their preferences in terms of privacy, access control, visibility radius, etc. Further meaningful extensions comprise mechanisms towards trust and reputation management to incentivize users to participate as well as to discourage malicious behavior. In Section~\ref{sec:roadmap}, we outline a roadmap towards real-world VLBS application, but full considerations of all involved aspects is beyond the scope of this paper.

\section{Data Collection and Generation}
\label{sec:setup}
We have considered two major cities in Ireland, Dublin and Galway, and collected data of places that feature a website, i.e., a virtual location. We then used data from Foursquare to simulate user movements between physical locations within these cities. We have studied these user movements to analyze the virtual locations they would visit during their movements in a city.

\subsection{Physical and Virtual Locations Data}
We have collected our datasets of virtual and physical locations within Dublin and Galway by using the \texttt{Google Places API}.\footnote{https://developers.google.com/places/} Since Google imposes a maximum result size on its search requests for places, the data collection process comprised the following four steps: (1)~We created a grid overlay for the areas of Galway and Dublin with a resolution of 100x100m, ensuring that search requests returned less than 200 results. (2)~For each grid point, we then issued \texttt{Radar Search} requests with a sufficiently large radius to retrieve all places around that grid point. (3)~Since \texttt{Radar Search} requests return only basic data about places, we then used \texttt{Place Details} requests to retrieve all information for all places found. (4)~Finally, we extracted all places that feature beside the geographic location also a virtual location, i.e., a website. As a result, we collected approximately 1,400 entries for Galway and 16,400 entries for Dublin 
with each entry featuring both a physical and virtual location. Section~\ref{sec:evaluation} provides full details. We made all data used in our evaluation available on an accompanying website.\footnote{http://vmusm02.deri.ie/vlimsy}

\subsection{Simulated User Movements}
\label{subsec:setup-simulated-user-movements}
To investigate how mobile users move through the virtual space, we simulated user movements across the cities of Dublin and Galway. On a large scale, studies have been carried out ~\cite{gonzalez2008understanding,sohn2006mobility} using cell tower data of mobile device users. More recently, human movement has been analyzed using the user check-in activities on location based social networks~\cite{noulas2012tale,nguyen2012using}. These works have shown that movements of any user occur within a specific geographical area with occasional movements outside the area. It has been demonstrated that many of the user movements have repeatability such as travelling to work place and users rarely travel between any random locations~\cite{oppenheim1995urban,pas1997recent}. The model advocating this is known as \emph{Activity-Based Travel Demand Modeling} and has been extensively used to model users' travel decisions.

\begin{SCfigure}
\centering
\includegraphics[width=0.50\textwidth]{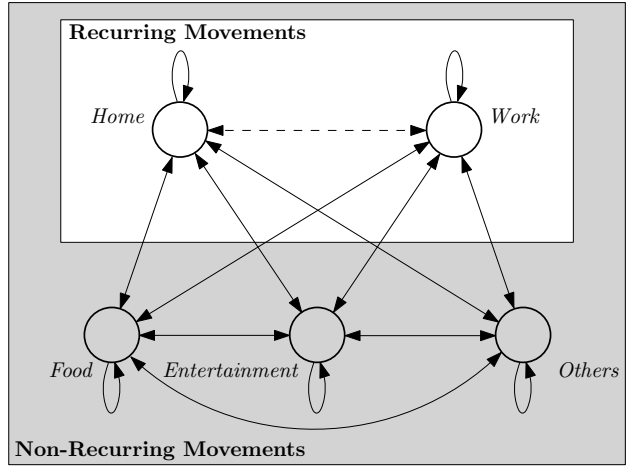}
\caption{User movements between various categories of places: Recurring movements refer to paths between \textit{Home} and \textit{Work} only; non-recurring movements consider all categories of places but excludes paths between \textit{Home} and \textit{Work}}
 \label{fig:user-movement}
\end{SCfigure}

We use these observations to simulate a user's start and end locations within a city. Specifically, we have used the total number of user check-ins, i.e., the number of visits, a place has on Foursquare to assess how much preference a user has to visit that place. Places in Foursquare belong to different categories such as 'restaurant', 'office', 'religious place', etc. In our simulation model, we have classified the place categories as $Home$, $Work$, $Food$, $Entertainment$ and $Others$. Similar schemes have been used in~\cite{noulas2011empirical, vasconcelos2012tips}. A user movement comprises of a user moving between places of different categories such as a user going from a movie theatre (\textit{Entertainment}) to a restaurant (\textit{Food}). We distinguish two types of user movements: \textit{recurring movements} between \textit{Home} and \textit{Work} represent the weekday routine of users; all other movements -- e.g., going to a movie theatre (\textit{Entertainment}) from the office (\textit{Work})
 -- we denote as \emph{non-recurring movements}. Figure~\ref{fig:user-movement} illustrates this, with \textit{Home} and \textit{Work} and the dashed edge reflecting recurring movements; non-recurring movements consider all five categories but exclude paths between \textit{Home} and \textit{Work}.

\begin{algorithm}[hbt]
\begin{algorithmic}[1]		
\State \text{select a home location} $l_{start}$ \text{as} $l_{start} \sim Uniform(1/h)$
\State \text{select a work place} $l_{end}$ \text{as follows}
\State $\pi_{W} \sim Dirichlet(\alpha_W)$
\State select $l_{end}$ as $l_{end} \sim Discrete(\pi_W)$
\State \text{return} ($l_{start}$, $l_{end}$) \text{and} ($l_{end}$, $l_{start}$)
\end{algorithmic}
\caption{Simulation of recurring movements}
\label{alg1}
\end{algorithm}
Algorithm ~\ref{alg1} describes the way samples of start and end locations are generated to simulate the recurring user movements. 
Here, $H$ and $W$ denote the set of all locations belonging to $Home$ and $Work$ categories respectively. A uniform sampling over $H$ with the probability of $1/h$, where $h=|H|$, ensures that every home location has equal probability of being the start location $l_{start}$ irrespective of the total number of check-ins at any home location (Line~1). We select the end location $l_{end}$ belonging to the \textit{Work} category based on the total number of check-ins. For this, we use the Dirichlet prior for a discrete distribution motivated by Bayesian Bootstrapping ~\cite{rubin1981bayesian}. We use this technique of smoothing the distributions as we have small datasets in terms of number of places per category and number of check-ins, compared to the number real-world locations and user visits. We first sample a distribution function $\pi_W$ from a Dirichlet distribution with the parameters $\alpha_W = (\alpha_1,\ldots,\alpha_w)$, where $w=|W|$ and $\alpha_i$ is the total number of user check-ins at the $i^{th}
$ work place (Line~3). This ensures that the discrete distribution sampled from $Dirichlet(\alpha_W)$ favors the selection of a work place as end location with a higher number of check-ins. Finally, we sample $l_{end}$ from a discrete distribution with the sample space $W$ and distribution function $\pi_W$ (Line~4).

\begin{algorithm}[hbt]
\begin{algorithmic}[1]		
\State \text{select a place category} $c_{start}$ \text{as} $c_{start} \sim Uniform(1/|C_{all}|)$
\If{$c_{start}$ \text{is home}}
	\State \text{select} $l_{start}$ \text{as} $l_{start} \sim Uniform(1/h)$	
\Else
	\State $\pi_{start} \sim Dirichlet(\alpha_{c_{start}}$)
	\State \text{select} $l_{start}$ \text{as} $l_{start} \sim Discrete(\pi_{start})$
\EndIf	

\State \text{select a place category} $c_{end}$ \text{as} $c_{end} \sim Discrete(w_{c_{start}})$
\If{$c_{end}$ \text{is home}}
	\State \text{select} $l_{end}$ \text{as} $l_{end} \sim Uniform(1/h)$	
\Else
	\State $\pi_{end} \sim Dirichlet(\alpha_{c_{end}}$)
	\State \text{select} $l_{end}$ \text{as} $l_{end} \sim Discrete(\pi_{end})$
\EndIf	
\State \text{return} ($l_{start}$, $l_{end}$)
\end{algorithmic}
\caption{Simulation of non-recurring movements}
\label{alg2}
\end{algorithm}
Algorithm~\ref{alg2} describes the way samples of $l_{start}$ and $l_{end}$ are chosen to simulate the non-recurring user movements. With $C_{all}$ being the set of all five categories of places, a uniform sampling is carried out on the sample space $C_{all}$ to obtain the category $c_{start}$ of any start location (Line~1), so that all categories are chosen equally. If $c_{start}$ is $Home$, we select any home location with equal probability (Line~3) to make sure that all home locations are well-represented in the simulations. If $c_{start}$ is not $Home$, we again use the Dirichlet prior for the discrete distribution to select $l_{start}$ to reflect the number of check-ins (Lines~5-6; cf. Algorithm~\ref{alg1}). The selection of $l_{end}$ comprises two steps: Firstly, we select a category $c_{end}$ based on the choice of $c_{start}$, favoring categories with many locations. And secondly, we select $l_{end}$ as location of category $c_{end}$, favoring locations with many check-ins. To select $c_{end}$, we 
use a stochastic transition matrix defined as:

$$
\bordermatrix{\text{}&Home&Work&Food&Entertainment&Others\cr
                 Home&\epsilon &  0  & w_{13} & w_{14} & w_{15} \cr
                 Work& 0  &  \epsilon & w_{23} & w_{24} & w_{25} \cr
                 Food& w_{31} & w_{32} & \epsilon & w_{34} & w_{35} \cr
                 Entertainment& w_{41}  &   w_{42}       & w_{43} & \epsilon & w_{45} \cr
                 Others& w_{51}  &   w_{52}       & w_{53} &  w_{54} & \epsilon  }
$$  
\\
\\
The transition probabilities between $Home$ and $Work$ are 0 here since we only consider non-recurring movements in Algorithm~\ref{alg2}. We calculate the transition probabilities between place categories as  $w_{ij} = \frac{N_j}{M_i} - \frac{\epsilon}{Z_i}$ with $i\neq j$. Here, $N_j$ is the number of distinct places belonging to $j^{th}$ category of the matrix. $M_i = \sum_{k}^{}N_k$ where $k$ is the any column index whose entry is not assigned with $\epsilon$ or 0 in the $i^{th}$ row. $Z_i$ is the number of place categories to which transition from the $i^{th}$ place category can be made. Hence, $Z_1 = Z_2 = 3$ and $Z_3 = Z_4 = Z_5 = 4$. $\epsilon$ denotes the self-transition probability of start and end location belonging to the same category. Setting the $\epsilon$ to a small value ensures that uncommon movements such as going from one \textit{Food} place another are not favored often in our simulations. Setting the $w_{ij}$ values based on the number of places belonging to various place categories 
ensures that the most visited place categories are favored in the simulation. Given $c_{start}$, we choose the end location category $c_{end}$ from a discrete distribution with the distribution function defined by the row vector $w_{c_{start}}$. Once we have select $c_{end}$, we sample an end location in a fashion similar to the selection of a start location (Lines~9-14).

Finally, for each selected start and end location, we used the \texttt{Google Directions API}\footnote{https://developers.google.com/maps/documentation/directions/} to obtain the paths between the two locations. Since we focus on walking users in our case study, we used \texttt{walking} as travel mode to get the directions via pedestrian paths and side-walks (where available). Note that with this method we consider direct paths locations without, e.g., loops. We discuss the effects of these characteristics on the evaluation result in Section~\ref{subsec:result-discussion}.

\section{Evaluation}
\label{sec:evaluation}
In this section, we present the results of our case study. The main question was whether merging the physical and virtual space into one space model provides a sufficient overlap to be of practical relevance. The input data for our analysis are the dataset of virtual locations and large samples of simulated user movements (see Section~\ref{sec:setup}). Table~\ref{tab:evaluation-parameters} lists and describes the parameters we considered within the analysis. The most important ones are the vicinity radius $r_v$ and the minimum visiting time $t_v^{min}$, together specifying what constitutes a visit of a user at a virtual location.
\begin{table}
 \centering
 \begin{tabular}[t]{|c|p{7cm}|}
  \hline
  $r_v$ & \textit{radius of vicinity} of virtual locations: minimum distance (in meter) between users and locations to be considered as visits of the users at locations. \\
  \hline
  $t_v^{min}$ & \textit{minimum visiting time}: minimum time (in seconds) a user has to spend in the vicinity radius of a virtual location to be considered as visit of users at locations. \\
  \hline
  $l_{max}$ & \textit{maximum path length}: maximum length in kilometer of simulated paths.\\
  \hline
 \end{tabular}
 \caption{List of evaluation parameters}
 \label{tab:evaluation-parameters}
\end{table}

\subsection{Simulated User Movements}
We crawled the details including the category and check-in activities for places in Dublin and Galway on Foursquare. Table~\ref{tab:place-details} shows the number of places for our five categories. We used this data to obtain the parameters of the various probability distributions in our simulation algorithms (see Section~\ref{subsec:setup-simulated-user-movements}).  
\begin{table}
 \centering
    \begin{tabular}{ | l | c | c | c | c | c | }
    \hline
     & \textit{Home} & \textit{Work} & \textit{Food}  & \textit{Entertainment} & \textit{Others} \\ \hline\hline
	Dublin & 4413 & 1280 & 1425  & 634 & 2269\\ \hline
	Galway & 417 & 228 & 275  & 156 & 442\\    
    \hline
    \end{tabular}
     \caption{Number of places belonging to different categories}
 \label{tab:place-details}
\end{table}
Table~\ref{tab:stationary-dist} shows the stationary distributions obtained for the five different place categories. These distributions are obtained by computing the long-term behavior of the Markov chain defined by the transition matrix defined in Section~\ref{subsec:setup-simulated-user-movements}. We set $\epsilon = 0.1$ as the self-transition probability parameter for the transition matrix. We found that the stationary distributions of user check-ins to places of the different categories are very similar for Dublin and Galway and also are similar to the result reported in the previous studies \cite{scellato2011exploiting,vasconcelos2012tips}.     

\begin{table}
 \centering
    \begin{tabular}{ | l | c | c | c | c | c | }
    \hline
     & \textit{Home} & \textit{Work} & \textit{Food}  & \textit{Entertainment} & \textit{Others} \\ \hline\hline
	Dublin & 0.184 &  0.158 &  0.232  &  0.322 &  0.104\\ \hline
	Galway & 0.106 &  0.181 &  0.241  &  0.323 &  0.149 \\    
    \hline
    \end{tabular}
  \caption{Stationary distributions of check-ins per category}
 \label{tab:stationary-dist}
\end{table}
 
\begin{SCfigure}
 \centering
 \includegraphics[width=0.65\textwidth]{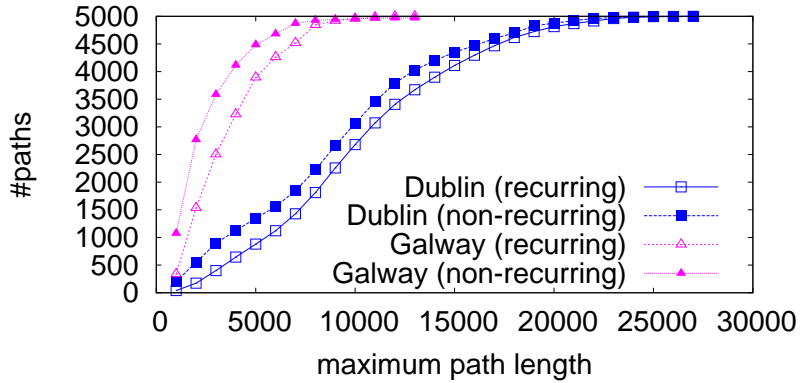}
\caption{Distribution of path length of the simulated recurring and non-recurring user movements in Dublin and Galway}
\label{fig:path-length-distribution-accumulated}
\end{SCfigure}

To obtain a reasonable number of user movements, we first generated 5,000 paths for recurring and non-recurring user movements according to our simulation model (see Section~\ref{subsec:setup-simulated-user-movements}) and the corresponding data for Dublin and Galway. Figure~\ref{fig:path-length-distribution-accumulated} shows the distribution of path lengths, i.e., the number of paths with lengths shorter than a given maximum path length, for each of the movement datasets. As expected, given the much larger size of the city, paths in Dublin are on average much longer than in Galway. Furthermore, popular locations for non-recurring movements (entertainment, food, others) are much more concentrated within the city of Galway. The figure also shows that paths for recurring movements are on average longer than for non-recurring movements. This is because home locations are typically outside the city center in residential areas which in turn feature not many work locations.

In the scope of this paper, we consider walking mobile users as main target group for VLIMSy. Naturally, paths with lengths of many kilometers do not represent meaningful walking paths. We therefore used for our evaluation the following approach: for experiments with a fixed maximum path length we set $l_{max} = 3$km; For experiments explicitly measuring the effect of $l_{max}$, we vary the maximum path length between 1 and 5 kilometers. Except for non-recurring movements in Dublin and $l_{max} = 1$km, each setting results in samples of at least 100 paths. Additional experiments showed that larger sample sizes do to not alter the overall results.

\subsection{Coverage \& Distribution of Locations}
\begin{figure}
 \centering
 \includegraphics[width=0.95\textwidth]{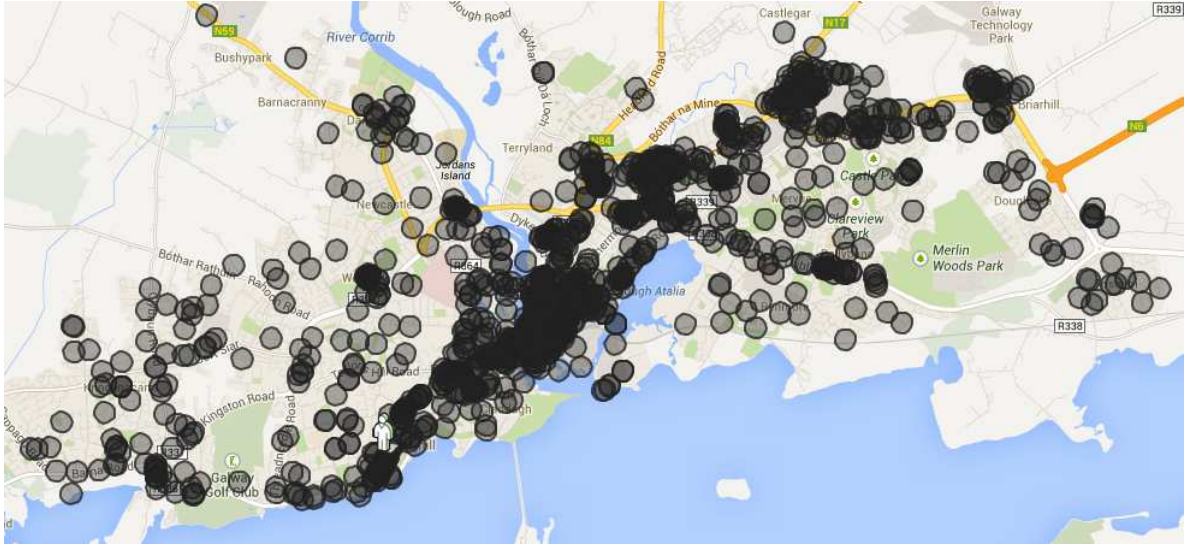}
\caption{Coverage of Galway City by virtual locations with vicinity radiuses $r_v=100$m}
\label{fig:virtual-locations-coverage-map}
\end{figure}

We now analyze the presence of virtual locations in urban areas. Table~\ref{tab:basic-data-stats} shows the population size, number of places and virtual locations (i.e., places that feature a website) we collected for both cities. While Dublin and Galway differ significantly regarding their population size, the number of places is roughly proportional to the size. Moreover, the number of places that feature a website is for both cities about 40\%. 
\begin{table}
 \centering
 \begin{tabular}[t]{|l|c|c|}
  \hline
  & \textbf{Dublin}	& \textbf{Galway} \\
  \hline\hline
  population (in 2011)	& 525,400	& 75,500 \\
  \hline
  \#places & 39,237  & 3,692 \\
  \hline
  \#virtual locations & 16,485 (42.0\%)  & 1455 (39.4\%) \\
  \hline
 \end{tabular}
 \caption{Basic statistics of collected data}
 \label{tab:basic-data-stats}
\end{table}
To give a first impression about the coverage and distribution of virtual locations, Figure~\ref{fig:virtual-locations-coverage-map} shows all locations within Galway City with a vicinity radius $r_v = 100$m. The figure already indicates that virtual locations are rather common and cover large portions of urban areas. Locations are numerous and spread across whole city, but not equally distributed -- this also holds for Dublin. Areas with a high density of locations are typically city centers, business parks or points of interests for tourists. 

\begin{SCfigure}
 \centering
 \includegraphics[width=0.6\textwidth]{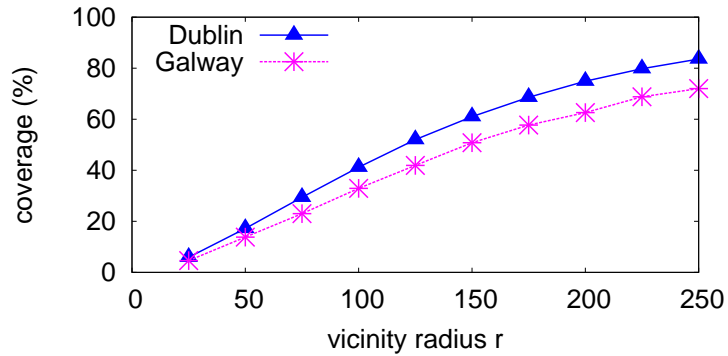}
 \caption{Coverage of virtual locations for Dublin and Galway and various vicinity radius $r_v$}
 \label{fig:virtual-locations-coverage}
\end{SCfigure}
\begin{SCfigure}
 \includegraphics[width=0.6\textwidth]{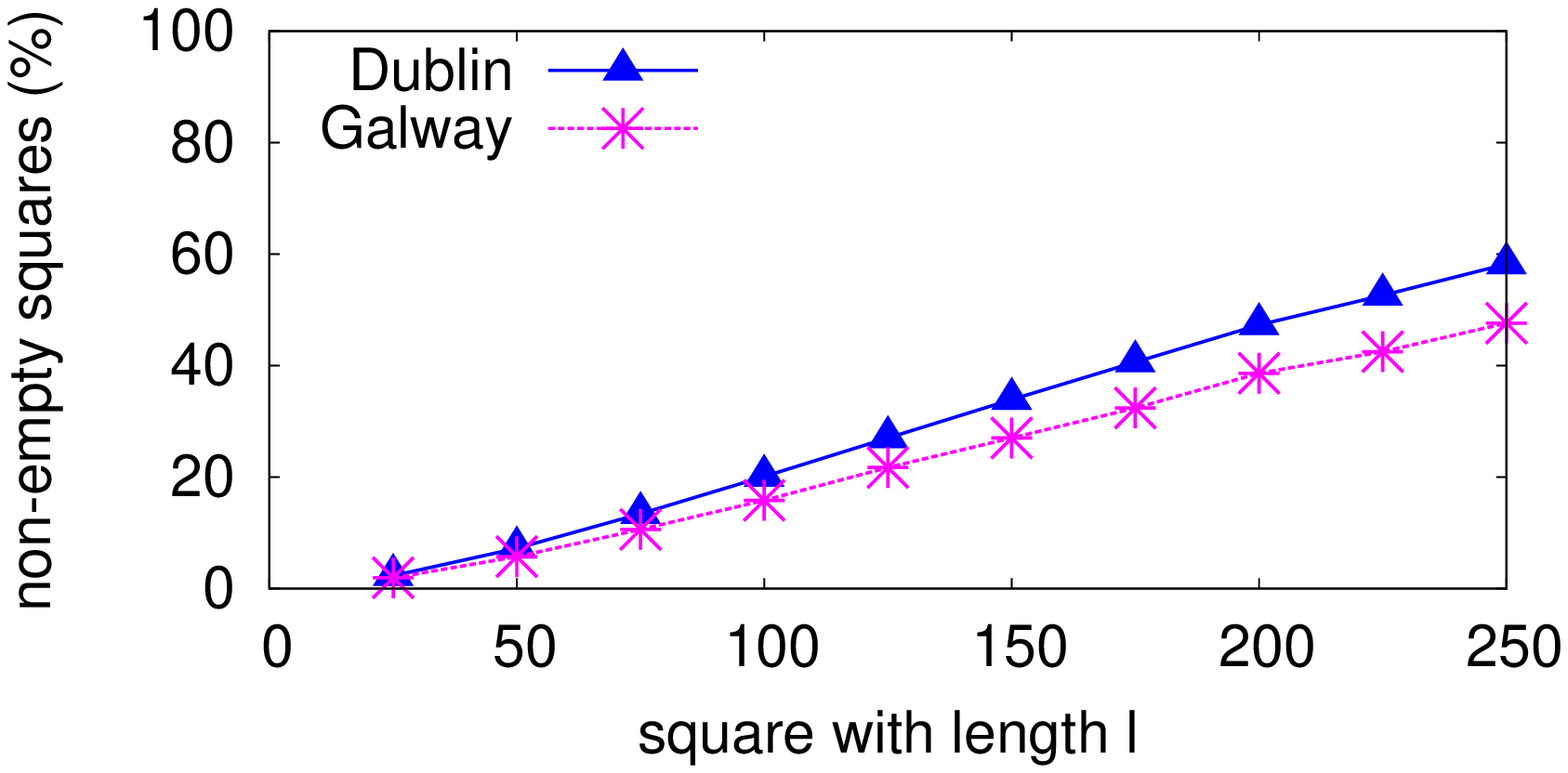}
 \caption{Ratio of squares across Dublin and Galway with one or more virtual locations for different square sizes}
\label{fig:virtual-locations-distribution-development}
\end{SCfigure}
\begin{SCfigure}
 \includegraphics[width=0.6\textwidth]{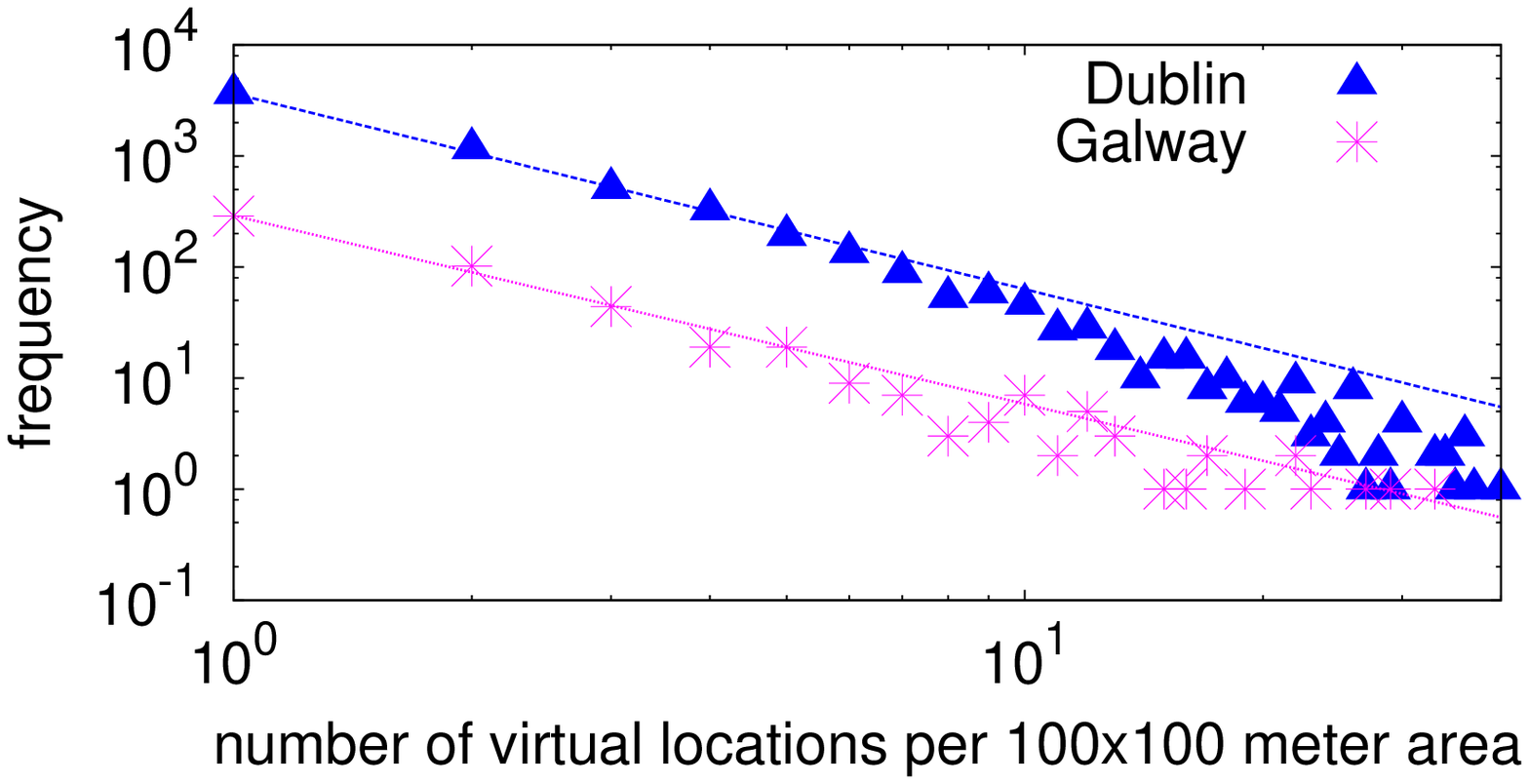}
 \caption{Distribution of non-empty squares across Dublin and Galway for different square sizes}
\label{fig:virtual-locations-distribution}
\end{SCfigure}

To get quantitative results, we first calculated the coverage in percent; see  Figure~\ref{fig:virtual-locations-coverage} for the results. Naturally, the coverage increases for larger vicinity radiuses, resulting in up to 83\% (72\%) coverage for Dublin (Galway) for $r = 250$m. In a second test, we looked in more detail into the distribution of the virtual locations. For this, we divided the areas of the cities into squares with side length $l$, with $l \in \{25, 50, 75, ..., 250\}$, and counted the number of virtual locations within each square. Figure~\ref{fig:virtual-locations-distribution-development} shows the ratio of non-empty squares, which naturally increases for larger squares. Empty squares typically cover city parks or purely residential areas. Figure~\ref{fig:virtual-locations-distribution} shows the distribution all non-empty squares for $l = 100$m. Not unexpectedly, the number of virtual locations per square and their respective frequency show a power-law relationship: While most squares 
contain only a small set of locations, few squares contain a very large number of virtual locations (city centers, business parks, and touristic points of interest). The power-law relationship prevails for all values of $l$, but of course with varying parameters describing the actual distributions.

These results confirm our initial assumption that virtual locations are very common, even omnipresent, in urban areas. Their wide distribution and coverage of Dublin and Galway already indicate that the probability of a mobile user being close to a virtual location at any point in time is very high. This is particularly true since we expect walking users to cross areas with a high density of virtual locations, such as the city center. We will show this in the following paragraphs using simulated user movements to get a picture how often and long mobile users visit virtual locations.

\begin{SCfigure}
 \centering
 \includegraphics[width=0.6\textwidth]{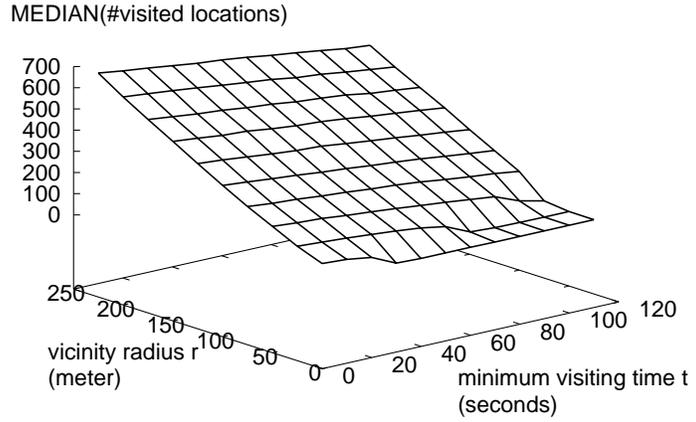}
 \caption{Average number of visited locations for different values of $r_v$ and $t_v^{min}$ and non-recurring movements in Dublin}
 \label{fig:radius-vs-visit-time-median-2000000-2005000-3000}
\end{SCfigure}
\begin{SCfigure}
 \includegraphics[width=0.6\textwidth]{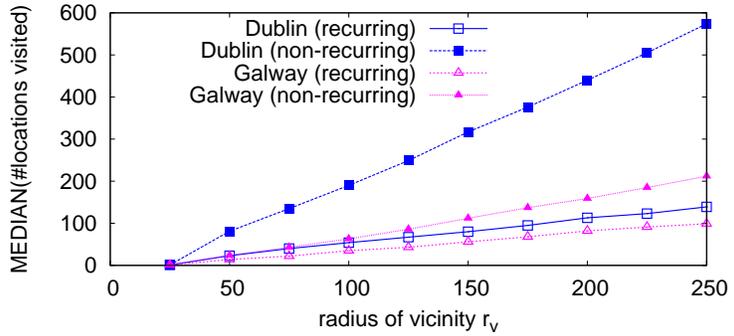}
 \caption{Average number of visited locations for each path datasets of Dublin and Galway ($t_v^{min} = 60\mathrm{s}$, $l_{max} = 3km$)}
\label{fig:radius-vs-visit-time-median-all-3000}
\end{SCfigure}
\begin{SCfigure}
 \includegraphics[width=0.6\textwidth]{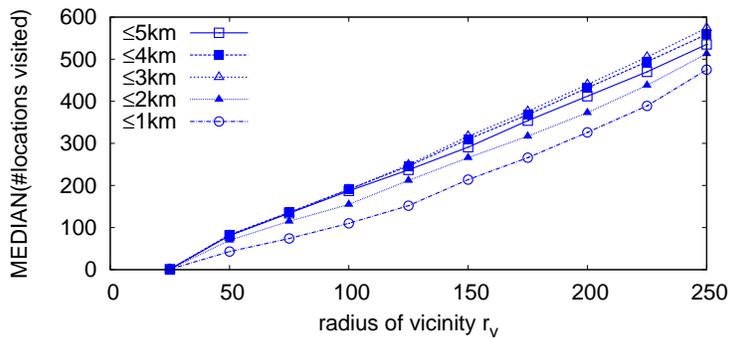}
 \caption{Average number of visited locations for different maximum path lengths $l_{max}$ ($t_v^{min} = 60\mathrm{s}$, non-recurring movements in Dublin)}
\label{fig:radius-vs-visit-time-median-all-length-dub-irregular}
\end{SCfigure} 

\subsection{Overlap of Physical and Virtual Space}
The main goal of our evaluation is to investigate the potential benefits of virtual location-based services. This translates to the question how the movements of mobile user actually overlap with virtual locations (together with a specified vicinity radius). Without a significant overlap, web and mobile users are very unlikely to ``meet'', rendering VLBS useless if this probability was too low.
\\
\\
\textbf{Average number of visited locations.} In first series of experiments, we investigated the number locations a mobile user is likely to cross while walking along path. Since the path lengths differ and hence the number of visited locations is skewed, we use the median to quantify the average number of visited locations.
Figure~\ref{fig:radius-vs-visit-time-median-2000000-2005000-3000} shows the results for the non-recurring movements in Dublin for various vicinity radiuses $r_v$ and minimum visiting times $t_v^{min}$ and for paths with a length of $\leq 3$km. The results for the other three path datasets are qualitatively almost identical and differ only in the absolute values. Most naturally, the smaller $r_v$ and the larger $t_v^{min}$ the less locations a user is visiting. The vicinity radius has the greater effect on the results due to the low speed (average walking speed). Overall, the results indicate that visiting many virtual locations while walking through a city is a very common phenomenon.

Next, we looked at the difference between the path datasets with otherwise identical parameter settings (Figure~\ref{fig:radius-vs-visit-time-median-all-3000}). As expected, the paths derived from non-recurring movements in Dublin cross the most virtual locations, since the paths typically pass through areas, e.g., the city center, that feature a very high density of locations. Recurring paths, on the other hand, often originate outside such areas, resulting in a significant lower number of visited locations. For Galway, the differences between recurring and non-recurring paths are also clearly visible. However, particularly for non-recurring paths, the results are much lower than for Dublin since the areas with a high density of virtual locations are much smaller. We therefore argue that VLBS are more likely to be successful in larger cities. Although the number of virtual locations is roughly proportional to the size of a city (cf. Table~\ref{tab:basic-data-stats}), large cities feature larger 
areas with a high density of virtual locations. This also includes that in such areas walking users are particularly common (e.g., city centers with or without pedestrian zones).

We also varied the maximum path length $l_{max}$ to see its effect on the average number of visited locations. Again, the results for the different path datasets differ only regarding the absolute values. As Figure~\ref{fig:radius-vs-visit-time-median-all-length-dub-irregular} shows, the value of $l_{max}$ affects the result only to a limited extend. This includes that the average number of visited locations slightly drops again for increasing maximum path lengths. Our explanation is that longer paths are more likely to originate from, pass or end in areas with a lower location density. In this sense, the maximum path length resulting in the highest average number of visited locations vaguely indicates the size of high-density areas. Note, however, that this assumption only holds in our current setting where we consider only direct paths between locations. If, for example, one considers only arbitrary paths confined to city center, we expect the number of individual visits -- this might also include repeated 
visits of the same virtual locations -- to be proportional to the maximum path length.
\\
\\
\textbf{Average number of parallel visits.} Depending on the density of virtual locations within an area, it is very likely that a mobile user is close the multiple locations at the same time. To quantify that expectation, we performed several experiments to measure the average number of parallel visits. Note that for each individual path the number of parallel visits changes over time and is typically rather skewed. We therefore calculated the median to represent the average number of parallel visits for each path. Since this average number also varies significantly between paths -- if not stated otherwise -- we eventually calculated the median of medians to report the average number of visits across our path datasets.

\begin{SCfigure}
 \centering
 \includegraphics[width=0.6\textwidth]{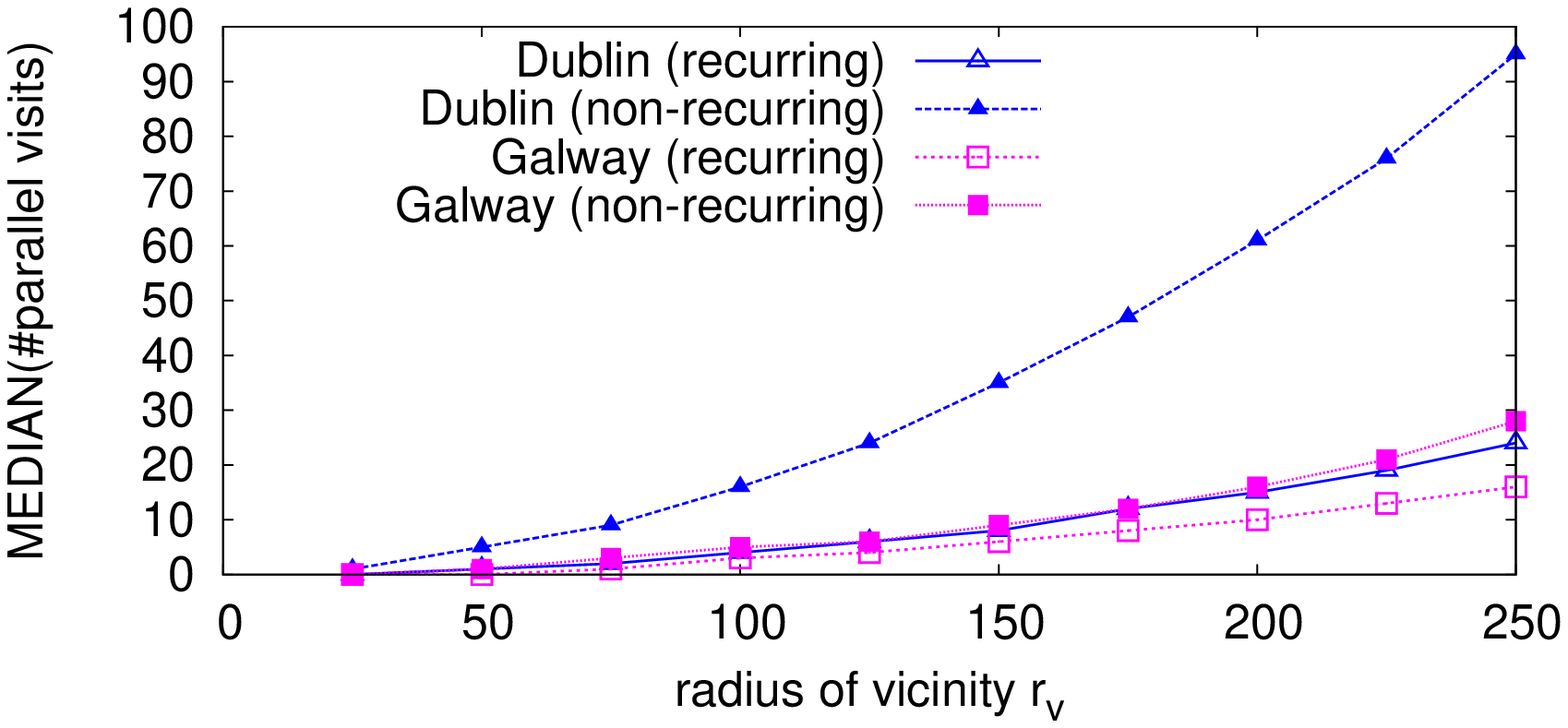}
 \caption{Average number of parallel visits for different graph dataset ($l_{max} = 3km$)}
 \label{fig:parallel-visits-all-datasets-3000}
\end{SCfigure}

\begin{SCfigure}
 \includegraphics[width=0.6\textwidth]{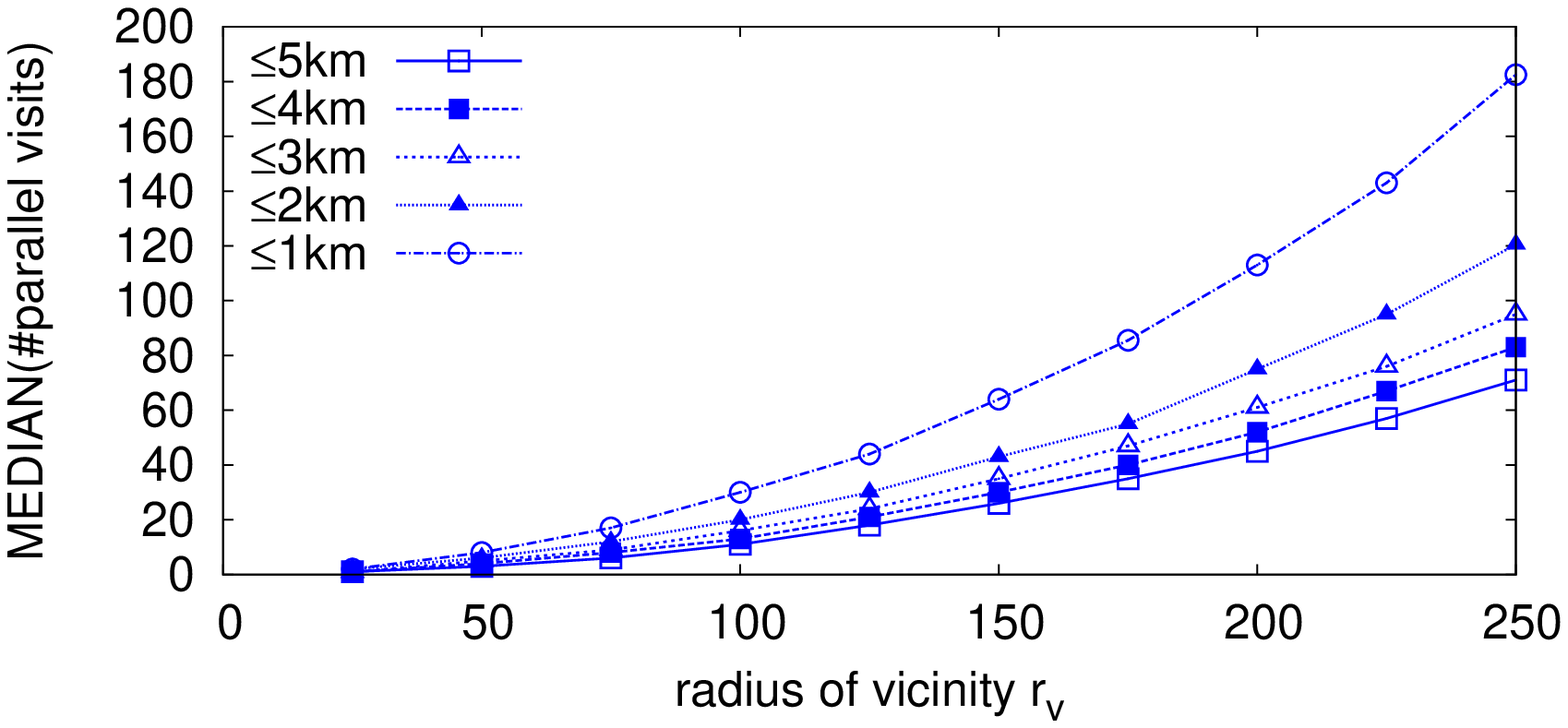}
 \caption{Average number of parallel visits for different maximum path lengths $l_{max}$ ($t_v^{min} = 60\mathrm{s}$, non-recurring movements in Dublin)}
\label{fig:parallel-visits-all-length-dub-irregular}
\end{SCfigure}

Figure~\ref{fig:parallel-visits-all-datasets-3000} shows the results for our path datasets. Not surprisingly, the number of parallel visits correlate with the results regarding the overall number of visited locations (cf., Figure~\ref{fig:radius-vs-visit-time-median-2000000-2005000-3000}). Even for moderate vicinity radiuses, the situation that a mobile user is close to multiple virtual locations at the same time is very common. For example, users following non-recurring paths in Dublin are, on average, close to 16 virtual locations. We then compared the results regarding the number of parallel visits for different maximum path lengths $l_{max}$. Figure~\ref{fig:parallel-visits-all-length-dub-irregular} shows -- by example for the non-recurring user movements in Dublin; the other datasets yield qualitatively similar results -- that for short maximum path lengths the number of parallel visits increases. This is, again, due to the higher probability of shorter paths to cross areas with high density of 
virtual locations. Being close to many virtual locations at the same time also increases the likelihood for web and mobile users to share the same location. 
\begin{SCfigure}
 \includegraphics[width=0.6\textwidth]{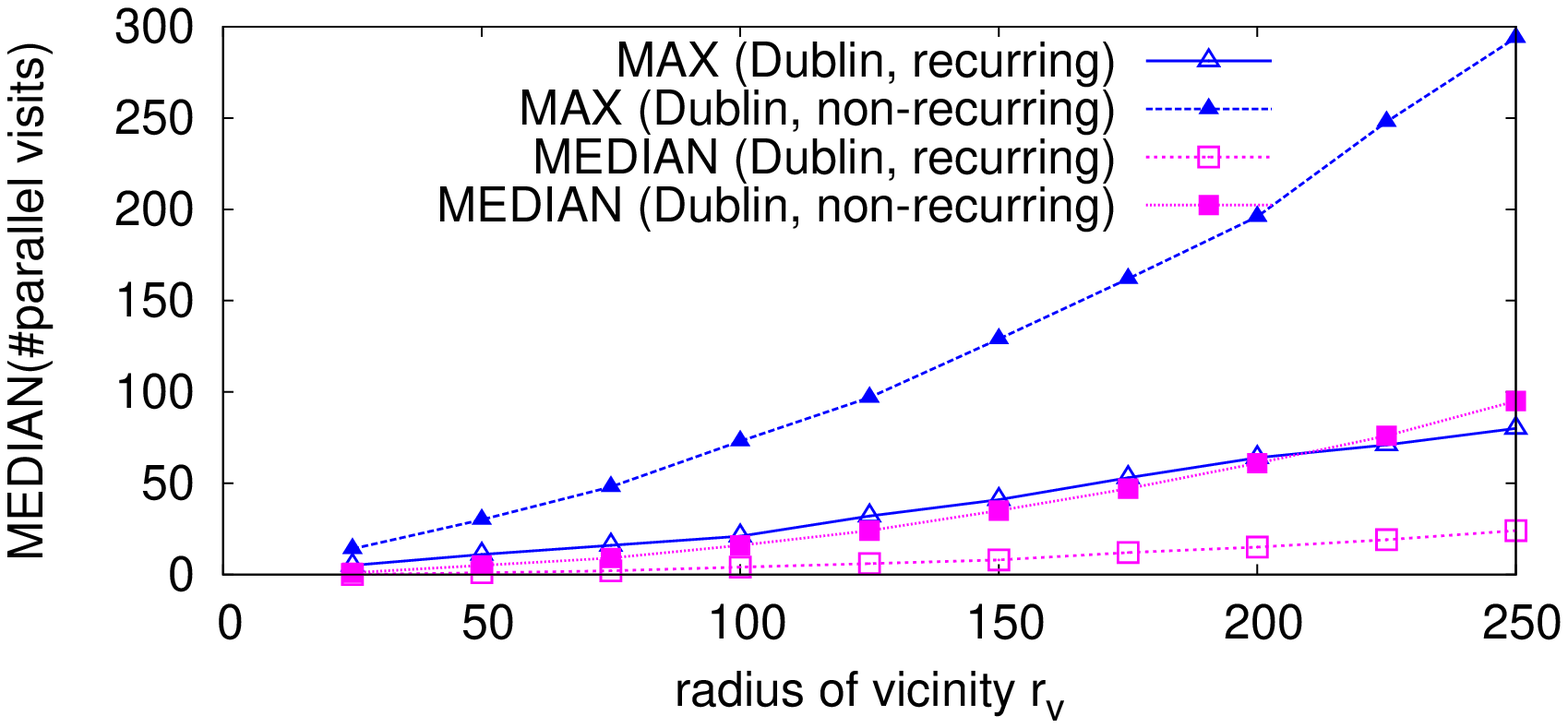}
 \caption{Average number of parallel visits: `Median of medians'' vs. ``median of maximums'' ($l_{max} = 3km$, Dublin)}
\label{fig:parallel-visits-median-vs-max-dub}
\end{SCfigure}  

To better see to how many virtual locations a user can be close at the same time, Figure~\ref{fig:parallel-visits-median-vs-max-dub} compares the median of medians results with the ones for median of maximums for the two Dublin path datasets. For the median of maximums, we first calculated the maximum number of parallel visited locations (instead of the median) for each path and then calculated the average over sets of paths using the median. Naturally, the average of maximums is significantly higher than the median of medians, resulting in up to several hundred parallel visits in parallel for non-recurring paths in Dublin. From an application perspective this means that that web and mobile user potentially find themselves in the presence of a large group of other users visiting the same locations at the same time. While this is, in general, a worthwhile situation, it also poses new challenges. Depending on the idea behind a VLBS, a fruitful communication or collaboration with others in parallel is limited 
to a manageable 
number of parallel interactions. In case of VLIMSy, for example, receiving many messages from many users in different contexts may be perceived as rather annoying and hence discourage users to participate. In Section~\ref{subsec:result-discussion} we discuss this issue on the design and implementation of VLBS.
\\
\\
\textbf{Accumulated visiting time.} In our last series of experiments, we evaluated the overlap between the physical and virtual spaces in terms of the average accumulated visiting time. This value represents the overall time a mobile user was close to virtual locations independent of if this visits were at the same time. For example, if a user was close to two locations for 1 minute at the same time, the accumulated visiting is 2 minutes. As such, the accumulated visiting time depends on the overall number of visited locations as well as the number of parallel visits. We regard it therefore as a combined measure to quantify the overlap of our user movements and the distribution of virtual locations.

\begin{SCfigure}
 \centering
 \includegraphics[width=0.6\textwidth]{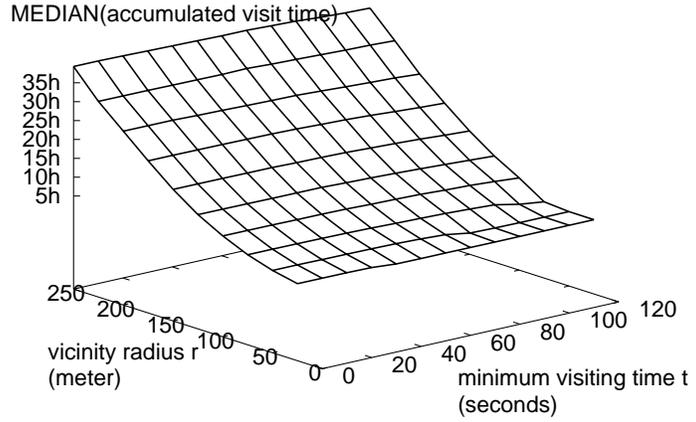}
 \caption{Accumulated visiting time for different values of $r_v$ and $t_v^{min}$ and non-recurring movements in Dublin}
 \label{fig:accumulated-time-radius-vs-time-median-3000}
\end{SCfigure}

\begin{SCfigure}
 \includegraphics[width=0.6\textwidth]{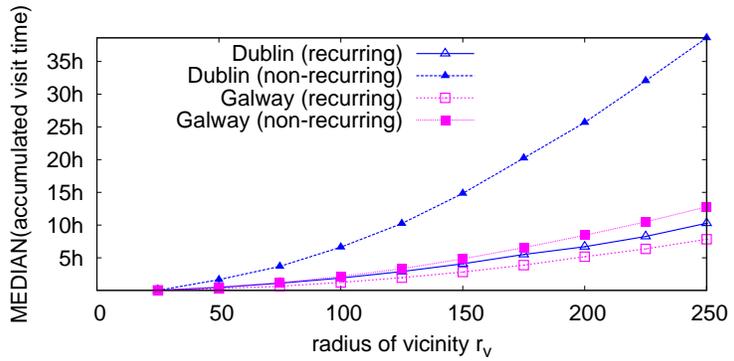}
 \caption{Accumulated visiting time for different different graph dataset ($l_{max} = 3km$)}
\label{fig:accumulated-time-all-datasets-3000}
\end{SCfigure}

We first calculated the accumulated times for varying vicinity radiuses $r_v$ and minimum visiting times $t_v^{min}$. Figure~\ref{fig:accumulated-time-radius-vs-time-median-3000} shows the results for the non-recurring movements in Dublin for paths with a length of less than 3km, where the accumulated times is the median over the accumulated visiting times of each path. Again, the effect of $t_v^{min}$ is less pronounced than the one of $r_v$ due to the slow travel speed of walking users. Most prominently, however, is the long accumulated visiting times of several hours, even for moderate vicinity radiuses. The reason for that is the typically high number of virtual locations a mobiles user is close-by at the same time. Figure~\ref{fig:accumulated-time-all-datasets-3000} shows the results for all path datasets and $l_{max} = 3\mathrm{km}$. As expected from previous results (cf. Figures~\ref{fig:radius-vs-visit-time-median-all-3000} and~\ref{fig:parallel-visits-all-datasets-3000}), the results lower for paths 
in 
Galway than for paths in Dublin. However, even for recurring movements in Galway, the average accumulated visiting time goes up to several hours. Thus, although mobile users are only present in the physical space for a rather short period of time -- not more than 36 minutes given a walking speed of 5km/h and $l_{max} = 3$km -- they are typically significantly longer present in the virtual space.

Finally, we investigated the effect of the maximum path length $l_{max}$ on the average accumulated visiting time. Again, as previous results have shown, longer maximum paths length do not result in longer visiting times since long paths tend to (partly) cross areas with a lower density of virtual locations. This is a result of our experiment setup using direct paths with a constant speed as user movements, and hence might differ for different use cases. In general, our setup yields the ``worst'' results in terms of the accumulated visiting time, since we, for example, do not consider any breaks during a walking path. Breaks, however, would be rather common when modeling, e.g., tourist who stop for shopping or eating. We elaborate on the effect of the chosen setup on the results in the following section.

\begin{SCfigure}
 \includegraphics[width=0.6\textwidth]{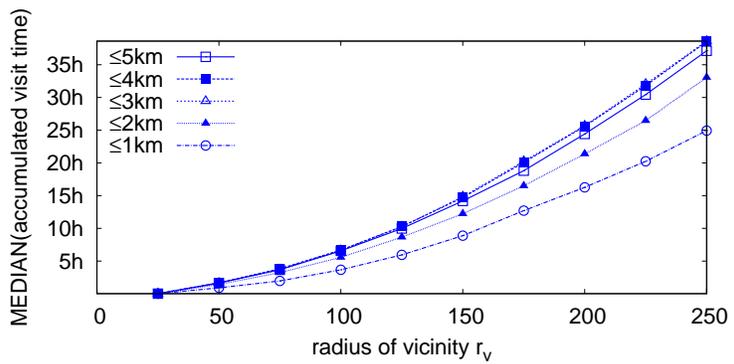}
 \caption{Accumulated visiting time or different maximum path lengths $l_{max}$ ($t_v^{min} = 60\mathrm{s}$, non-recurring movements in Dublin)}
\label{fig:accumulated-time-all-length-dub-irregular}
\end{SCfigure}  

\subsection{Discussion of Results}
\label{subsec:result-discussion}
Our consistent results over our chosen scenario make a clear case for the notion of virtual location-based services and related applications. They also provide first insights into the requirements for such services and applications. We summarize our results in the following to back up these claims.
\\
\\
\textbf{Potential benefits of VLBS.} 
To be useful, VLBS require that the physical space, i.e., geographic locations, and the virtual space, i.e., websites, overlap sufficiently. Our results clearly show that in urban areas mobile users passing virtual locations is the normal rather than a special case. Moreover, areas like city centers or commercial districts feature such a high density of virtual locations that users are close to a large number of locations at any time. The main reason for this is the large number of locations that feature a physical as well as virtual location. Thus, we deem the virtual presence of mobile users of practical importance for the design and development of new kinds of location-based services bridging the physical and virtual world. Although our results are already very promising, we expect a more significant overlap in real-world deployments. Firstly, our observations show that there are more virtual locations within Galway and Dublin that are not returned by \texttt{Google Places}, for example, small businesses 
or shops. As such, we deem the number of virtual locations in our datasets as lower bounds. Secondly, we considered for our evaluation only the use case where mobile users walk on a direct path from A to B. Imagine, however, a typical tourist strolling through the city center of Dublin: The movement is no longer a direct path and might even cross itself multiple times, the walking speed is slow and includes breaks for shopping or eating. This further increases the overlap between the virtual and physical world, particularly regarding the time spent close-by virtual locations.
\\
\\
\textbf{Impact on implementations of (virtual) location-based services.}
In areas with a high density of virtual locations such as city centers, the probability that a mobile user is present at a large number of virtual locations at the same time is very high. Depending on the number of already present users this can lead to an unmanageable number of parallel encounters, and therefore negatively affect the user experience. Thus, suitable mechanisms limiting a user's presence to a reasonable number of parallel virtual locations are required. Firstly, one can apply filter and ranking techniques to limit the number of virtual locations a mobile user is considered to be visiting. Meaningful techniques consider the different absolute distances or more sophisticated parameters (e.g., how often a user is close a particular location) to determine the top-k list of locations a user is actually deemed present. A further approach is to dynamically adapt important system parameters such as the vicinity radius $r_v$ or the minimum visiting time $t_v^{min}$. For example, it is conceivable to 
adjust these values for different locations independently, according to the current density of virtual locations or the overall number of already present users.

\section{A Roadmap Towards VLBS}
\label{sec:roadmap}
In the following, we discuss the effects of our evaluation results on the development of new (virtual) location-services and where we see interesting follow-up research questions.
\\
\\
\textbf{Privacy concerns.}
Like for ``traditional'' location-based services, privacy is also a relevant issue for VLBS. For example, a mobile user does not want to be associated with the website of an erotic shop. Given the accuracy of GPS, it is often hard to distinguish if a mobile user is actually in a specific shop or in an adjacent one. Existing approaches to preserve users' privacy essentially aim to not disclose the exact location but rather an estimate (see, e.g., \cite{Krumm09Survey,Shin12PrivacyProtection} for an comprehensive overview). The main challenge here is identifying a meaningful trade-off between the level of privacy and the quality of the provided service. It needs to be investigated if and how existing techniques can be applied to VLBS. On the other hand, VLBS can leverage from related areas such as online social networks. Here, for example, users can formulate or set policies to express to which groups of people (family, friends, etc.) which visited virtual locations they want to disclose.
\\
\\
\textbf{Setting incentives.} VLBS applications such as VLIMSy rely on the contribution of its users. In general, users are willing to contribute if their perceived benefit significantly outweighs their perceived costs. The perceived benefit derives, most importantly, from the added value the services offers to users. Additional techniques to stimulate cooperative behavior might include incentive mechanisms such as reputation systems or scoring schemes. The perceived costs typically range from the effort of users to contribute (e.g., clicking a rating, adding tags, writing a review, etc.) to their perceived privacy risks. Again, we argue that the integration of existing social network relationships since users are more likely willing to interact with friends. Note also that there is an inherent incentive for users to participate and contribute. For example, a mobile user can be a shop owner and as such has a straightforward interest to be in touch with visitors of his/her shop's website.
\\
\\
\textbf{Extended concepts.}
To ease the presentation of our results, we made some simplifying assumptions regarding our evaluation. Real-world (virtual) location-based services may benefit from extended and additional concepts to provide better services to users. Some meaningful extensions are:

\textit{(1) Location predictions.} So far, we have limited ourselves to results that refer to the current location of users. There are, however, different approaches conceivable to predict probable locations in the (near) future. This can be done using time series analysis of particularly repeated movements such as people's daily trip to work and back home. Knowledge about recurring movements allows indicating one's presence on a virtual location before the user is indeed close-by. Other approaches might consider the current location, speed, and direction of a mobile user to predict virtual locations the user is likely to be present in the future.

\textit{(2) Locations with spatial extent.} We considered only ``single point'' locations (e.g., hotels, pubs, shops) with no spatial extent. This includes that we currently store locations like parks, golf courses or nature reserves, using a single geo coordinate to represent their physical location. Associating the whole area of such locations to a virtual location is an intuitive next step. This may also include the definition of regions depending on a specific application scenario. For example, one can divide the city limits of Galway into the inner city, its suburbs, industrial areas, and so on.

\textit{(3) Semantic virtual locations.} Our current knowledge base contains mainly locations that ``naturally'' feature both a physical and virtual location, such as hotels, shops, restaurants, companies etc., all featuring their own website. However, there are all kinds of connections between physical and virtual locations conceivable. For example, one can connect a news article about Dublin to the the area of Dublin. Or one can connect a \textsc{YouTube} video about pubs in Dublin to the respective physical locations of the mentioned pubs. The spectrum of use cases that benefit from such connections is very broad. In this respect, our current knowledge base contains just the minimum of virtual locations. Hence, all our result regarding the cumulated visiting time, the number of parallel visits, and others, are only a lower bound, i.e., in many settings the achievable results will be better.

\textit{(4) Exploiting similarities between locations.} For our analysis, we considered all virtual location independent from each other. However, it might also be meaningful to group multiple virtual locations based on their similarities. For example, it might be more interesting if users are close to virtual locations associated with pubs than close to a virtual location of a specific pub. In this case, the similarity derives the same type of physical location (here: \textit{pub}). For different location-based services, different meaningful ways to derive such similarities are conceivable.

\section{Conclusions}
\label{sec:conclusions}
Merging virtual locations (i.e., webpages or sites) with points of interest in the real world, opens new opportunities for the design and development of novel virtual location-based services. Mobile users can get additional information about websites associated with nearby locations provided on their devices. Users at home browsing a website can get in contact with people on-site, i.e., users that are close to the location connected to the visited website. In this paper, we showed that such approaches promise to have real practical impact. Firstly, the number points of interest that feature both a physical and virtual location is very high. In our two datasets, virtual locations are spread across the cities of Dublin and Galway. Both cities also feature areas such as the city center or commercial districts with a very high number of virtual locations. As a result, even for moderate vicinity radiuses, virtual locations cover very large portions of urban areas. And secondly, using simulated tracks describing 
different categories of user movements, we demonstrated how mobile users move through the virtual space by being close to virtual locations. All our results consistently show that the situation where users are nearby such locations is rather the normal than the exceptional case. Moreover, as a result of the distribution of virtual locations, users are often close to many of them at the same time.

Summing up, our results show that there is a significant overlap between the physical and virtual space, which in turn promotes the practical relevance and potential benefits of VLBS. Furthermore, the results also serve as input for ideas and the design and implementation of such services. Building on our evaluation results, we presented a roadmap towards real-world VLBS. We firstly outlined immediate challenges, notably addressing privacy concerns and incentivizing users to contribute. We argue that not only existing techniques from traditional location-based services but also mechanisms from other related fields, such as online social networks, are worth investigating. Finally, we discussed conceivable extensions to our current approaches -- such as the consideration of locations with a spatial extent, or the association of physical and virtual locations based on a shared context -- that allow for the development new types of services with further additional benefits to users.

\bibliographystyle{abbrv}
\bibliography{literature}

\begin{thebibliography}{10}

\bibitem{backstrom2010find}
L.~Backstrom, E.~Sun, and C.~Marlow.
\newblock Find me if you can: improving geographical prediction with social and
  spatial proximity.
\newblock In {\em Proceedings of the 19th international conference on World
  wide web}, pages 61--70. ACM, 2010.

\bibitem{Bellavista08LBS}
P.~Bellavista, A.~Kupper, and S.~Helal.
\newblock {Location-Based Services: Back to the Future}.
\newblock {\em Pervasive Computing, IEEE}, 7(2):85--89, 2008.

\bibitem{Brush10ExploringEndUser}
A.~B. Brush, J.~Krumm, and J.~Scott.
\newblock {Exploring End User Preferences for Location Obfuscation,
  Location-Based services, and the Value of Location}.
\newblock In {\em Proceedings of the 12th ACM international conference on
  Ubiquitous computing}, Ubicomp '10, pages 95--104. ACM, 2010.

\bibitem{Chang07UserStudyOnAdoption}
S.~Chang, Y.-J. Hsieh, T.-R. Lee, C.-K. Liao, and S.-T. Wang.
\newblock {A User Study on the Adoption of Location Based Services}.
\newblock In K.-C. Chang, W.~Wang, L.~Chen, C.~Ellis, C.-H. Hsu, A.~Tsoi, and
  H.~Wang, editors, {\em Advances in Web and Network Technologies, and
  Information Management}, volume 4537 of {\em Lecture Notes in Computer
  Science}, pages 276--286. Springer, 2007.

\bibitem{cho2011friendship}
E.~Cho, S.~A. Myers, and J.~Leskovec.
\newblock Friendship and mobility: user movement in location-based social
  networks.
\newblock In {\em Proceedings of the 17th ACM SIGKDD international conference
  on Knowledge discovery and data mining}, pages 1082--1090. ACM, 2011.

\bibitem{Roza03OverviewOfLBS}
T.~D'Roza and G.~Bilchev.
\newblock {An Overview of Location-Based Services}.
\newblock {\em BT Technology Journal}, 21(1):20--27, 2003.

\bibitem{gonzalez2008understanding}
M.~C. Gonzalez, C.~A. Hidalgo, and A.-L. Barabasi.
\newblock Understanding individual human mobility patterns.
\newblock {\em Nature}, 453(7196):779--782, 2008.

\bibitem{Gu09SurveyOfIndoor}
Y.~Gu, A.~Lo, and I.~Niemegeers.
\newblock {A Survey of Indoor Positioning Systems for Wireless Personal
  Networks}.
\newblock {\em Communications Surveys Tutorials, IEEE}, 11(1):13--32, 2009.

\bibitem{Junglas05ExperimentalInvestigation}
I.~Junglas.
\newblock {An Experimental Investigation of Location-Based Services}.
\newblock In {\em System Sciences, 2005. HICSS '05. Proceedings of the 38th
  Annual Hawaii International Conference on}, pages 85a--85a, 2005.

\bibitem{Krumm09Survey}
J.~Krumm.
\newblock {A Survey of Computational Location Privacy}.
\newblock {\em Personal and Ubiquitous Computing}, 13(6):391--399, 2009.

\bibitem{Lassabe09IndoorPositioning}
F.~Lassabe, P.~Canalda, P.~Chatonnay, and F.~Spies.
\newblock {Indoor Wi-Fi positioning: Techniques and Systems}.
\newblock {\em Annals of Telecommunications}, 64(9-10):651--664, 2009.

\bibitem{Leshed08CoScripter}
G.~Leshed, E.~M. Haber, T.~Matthews, and T.~Lau.
\newblock {CoScripter: Automating \& Sharing How-to Knowledge in the
  Enterprise}.
\newblock CHI '08, pages 1719--1728, New York, NY, USA, 2008. ACM.

\bibitem{Liu07SurveyOfWireless}
H.~Liu, H.~Darabi, P.~Banerjee, and J.~Liu.
\newblock {Survey of Wireless Indoor Positioning Techniques and Systems}.
\newblock {\em Systems, Man, and Cybernetics, Part C: Applications and Reviews,
  IEEE Transactions on}, 37(6):1067--1080, 2007.

\bibitem{Morris07SearchTogether}
M.~R. Morris and E.~Horvitz.
\newblock {SearchTogether: An Interface for Collaborative Web Search}.
\newblock In {\em Proceedings of the 20th Annual ACM Symposium on User
  Interface Software and Technology}, UIST '07, pages 3--12, New York, NY, USA,
  2007. ACM.

\bibitem{nguyen2012using}
T.~Nguyen and B.~K. Szymanski.
\newblock Using location-based social networks to validate human mobility and
  relationships models.
\newblock In {\em Advances in Social Networks Analysis and Mining (ASONAM),
  2012 IEEE/ACM International Conference on}, pages 1215--1221. IEEE, 2012.

\bibitem{noulas2012tale}
A.~Noulas, S.~Scellato, R.~Lambiotte, M.~Pontil, and C.~Mascolo.
\newblock A tale of many cities: universal patterns in human urban mobility.
\newblock {\em PloS one}, 7(5):e37027, 2012.

\bibitem{noulas2011empirical}
A.~Noulas, S.~Scellato, C.~Mascolo, and M.~Pontil.
\newblock An empirical study of geographic user activity patterns in
  foursquare.
\newblock {\em ICWSM}, 11:70--573, 2011.

\bibitem{oppenheim1995urban}
N.~Oppenheim et~al.
\newblock {\em Urban travel demand modeling: from individual choices to general
  equilibrium.}
\newblock John Wiley and Sons, 1995.

\bibitem{pas1997recent}
E.~I. Pas.
\newblock Recent advances in activity-based travel demand modeling.
\newblock In {\em Activity-Based Travel Forecasting Conference}, 1997.

\bibitem{rubin1981bayesian}
D.~B. Rubin.
\newblock The bayesian bootstrap.
\newblock {\em The annals of statistics}, 9(1):130--134, 1981.

\bibitem{scellato2011socio}
S.~Scellato, A.~Noulas, R.~Lambiotte, and C.~Mascolo.
\newblock Socio-spatial properties of online location-based social networks.
\newblock {\em ICWSM}, 11:329--336, 2011.

\bibitem{scellato2011exploiting}
S.~Scellato, A.~Noulas, and C.~Mascolo.
\newblock Exploiting place features in link prediction on location-based social
  networks.
\newblock In {\em Proceedings of the 17th ACM SIGKDD international conference
  on Knowledge discovery and data mining}, pages 1046--1054. ACM, 2011.

\bibitem{Shin12PrivacyProtection}
K.~Shin, X.~Ju, Z.~Chen, and X.~Hu.
\newblock {Privacy Protection for Users of Location-Based Services}.
\newblock {\em Wireless Communications, IEEE}, 19(1):30--39, 2012.

\bibitem{sohn2006mobility}
T.~Sohn, A.~Varshavsky, A.~LaMarca, M.~Y. Chen, T.~Choudhury, I.~Smith,
  S.~Consolvo, J.~Hightower, W.~G. Griswold, and E.~De~Lara.
\newblock Mobility detection using everyday gsm traces.
\newblock In {\em UbiComp 2006: Ubiquitous Computing}, pages 212--224.
  Springer, 2006.

\bibitem{Toch10EmpericalModels}
E.~Toch, J.~Cranshaw, P.~H. Drielsma, J.~Y. Tsai, P.~G. Kelley, J.~Springfield,
  L.~Cranor, J.~Hong, and N.~Sadeh.
\newblock {Empirical Models of Privacy in Location Sharing}.
\newblock In {\em Proceedings of the 12th ACM international conference on
  Ubiquitous computing}, Ubicomp '10, pages 129--138. ACM, 2010.

\bibitem{vasconcelos2012tips}
M.~A. Vasconcelos, S.~Ricci, J.~Almeida, F.~Benevenuto, and V.~Almeida.
\newblock Tips, dones and todos: uncovering user profiles in foursquare.
\newblock In {\em Proceedings of the fifth ACM international conference on Web
  search and data mining}, pages 653--662. ACM, 2012.

\bibitem{vdw11COBS}
C.~von~der Weth and A.~Datta.
\newblock {COBS: Realizing Decentralized Infrastructure for Collaborative
  Browsing and Search}.
\newblock In {\em 2011 IEEE International Conference on Advanced Information
  Networking and Applications}, pages 617--624. IEEE, Mar. 2011.

\bibitem{vdw11FAST}
C.~von~der Weth and A.~Datta.
\newblock {FAST: Friends Augmented Search Techniques - System Design \{\&\}
  Data-Management Issues}.
\newblock In {\em Proceedings of the 2011 IEEE/WIC/ACM International
  Conferences on Web Intelligence and Intelligent Agent Technology - Volume
  01}, WI-IAT '11, pages 356--363, Washington, DC, USA, 2011. IEEE Computer
  Society.

\bibitem{vdw10COBSdemo}
C.~von~der Weth, A.~Datta, and S.~Ang.
\newblock {COBS: A tool for collaborative browsing and search on the web}.
\newblock In {\em 2010 IEEE International Conference on Multimedia and Expo},
  pages 272--273. IEEE, July 2010.

\bibitem{vdw11VirtualPresence}
C.~von~der Weth and M.~Hauswirth.
\newblock {Finding Information Through Integrated Ad-Hoc Socializing in the
  Virtual and Physical World}.
\newblock In {\em Proceedings of the 2013 IEEE/WIC/ACM International
  Conferences on Web Intelligence and Intelligent Agent Technology (to
  appear)}, WI-IAT '13, Atlanta, USA, 2013. IEEE Computer Society.

\bibitem{wang2011human}
D.~Wang, D.~Pedreschi, C.~Song, F.~Giannotti, and A.-L. Barabasi.
\newblock Human mobility, social ties, and link prediction.
\newblock In {\em Proceedings of the 17th ACM SIGKDD international conference
  on Knowledge discovery and data mining}, pages 1100--1108. ACM, 2011.

\bibitem{Wiltse09PlayByPlay}
H.~Wiltse and J.~Nichols.
\newblock {PlayByPlay: Collaborative Web Browsing for Desktop and Mobile
  Devices}.
\newblock In {\em Proceedings of the 27th International Conference on Human
  Factors in Computing Systems}, CHI '09, pages 1781--1790, New York, NY, USA,
  2009. ACM.

\bibitem{zheng2010geolife}
Y.~Zheng, X.~Xie, and W.-Y. Ma.
\newblock Geolife: A collaborative social networking service among user,
  location and trajectory.
\newblock {\em IEEE Data Eng. Bull.}, 33(2):32--39, 2010.

\end{thebibliography}

\end{document}